\documentclass{cs20proc}

\usepackage{kantlipsum}

\editors{S. J. Wolk}
\publisher{Zenodo}
\conference{The 20th Cambridge Workshop on Cool Stars, Stellar Systems, and the Sun}
\conferencedate{2018}

\title{Activity variation driven by flux emergence and transport on Sun-like stars}
\author{Emre I\c{s}{\i}k,$^{1,2}$ 
        Sami K. Solanki,$^{1,3}$
        Natalie A. Krivova,$^{1}$        
        Alexander I. Shapiro,$^{1}$}

\affiliation{$^{1}$ Max-Planck-Institut f\"ur Sonnensystemforschung, 
Justus-von-Liebig-Weg 3, 37077, G\"ottingen, Germany \\
$^{2}$ Feza G\"ursey Center for Physics and Mathematics, Bo\u{g}azi\c{c}i 
University, 34680 Kuleli, Istanbul, Turkey \\
$^{3}$ School of Space Research, Kyung Hee University, Yongin, Gyeonggi-Do, 
446-701, Republic of Korea}

\shorttitle{Flux emergence and transport on Sun-like stars}
\shortauthors{Emre I\c{s}{\i}k et al.}

\abs{In G dwarfs, the surface distribution, coverage and lifetimes of starspots deviate from solar-like patterns as the rotation rate increases. We set up a numerical platform which includes the large-scale rotational and surface flow effects, aiming to simulate evolving surface patterns over an activity cycle for up to 8 times the solar rotation and flux emergence rates. At the base of the convection zone, we assume a solar projected butterfly diagram. We then follow the rotationally distorted trajectories of rising thin flux tubes to obtain latitudes and tilt angles. Using them as source distributions, we run a surface flux transport model with solar parameters. Our model predicts surface distributions of the signed radial fields and the starspots that qualitatively agree with observations.}

\begin{document}

\maketitle

\section{Introduction}
Sun-like stars exhibit magnetically induced brightness and spectropolarimetric 
variability in a broad range of timescales. Rotational variations are induced by 
starspots whereas annual to decadal variations are potentially driven by activity 
cycles and/or by long-term effects of short-term stochasticity. Physical models of 
the surface 
distribution of magnetic activity on Sun-like stars can help us gain insight 
to better characterise physical processes involved. They can also be used 
in forward-modelling of brightness variations and compare with space-borne 
photometry. 

We developed a modelling platform to simulate large-scale magnetic fields 
over the stellar surface and in time \citep{isik18}. Here we present a summary 
of the method and a selection of simulation results for a set of rotation rates 
and activity levels. 

\section{Emergence of magnetic flux}
We take a Sun-like butterfly diagram of flux eruptions to represent initial latitudes 
of thin flux tubes and follow their buoyant rise up to the surface, to obtain 
the emergence latitudes and tilt angles at the surface. 
We follow the procedure outlined below, to set up the pattern of magnetic 
flux eruptions from the base of the convection zone. 
\begin{enumerate}
\item Magnetic flux eruptions from the base of the convection zone follow a 
statistical numerical model of the activity cycle at the solar surface \citep{jcss11a}. 
To simulate the rotation-activity relation, we set the flux emergence frequency 
(in solar units) equal to the stellar rotation rate (in solar units). 
\item Emergence latitude and tilt angle for a given eruption are calculated from 
numerical simulations of thin flux tubes rising through a non-local mixing-length 
solar convection zone stratification. 
\item The initial field strength of a flux tube is set by a function of the initial latitude. 
This 
function is determined by the linear stability condition for mechanical equilibrium of 
a thin flux ring in the convective overshoot region, for given stellar rotation rate. 
We take a linear growth time of 50 days for 
any flux tube. In this way, the field strength is close to the critical strength for the 
onset of magnetic buoyancy 
instability. We assume that the latitudinal and radial differential rotation amplitudes 
($\Delta\Omega)$ do not change for faster rotation. 
\item The nonlinear flux-tube simulations are stopped when the thin-flux-tube 
condition is violated, i.e., at about $0.98R_\odot$. The emergence latitude is 
recorded at this point. The tilt angle is calculated from the coordinates of the 
legs of the flux loop at $0.97R_\odot$. 
\end{enumerate}

\section{Transport of surface magnetic flux}
We take the emergence latitudes and tilt angles resulting from the procedure 
described above into a surface flux transport (SFT) model \citep[see also][]{isik11}. 
On top of the emergence model, 
we modify the emergence latitudes and longitudes in a probabilistic way, 
so as to allow for nesting of activity. In the current work, we defined only 
a single activity cycle and we did not impose random scatter about the tilt angles, 
which we derived from rising flux tubes. 

\begin{figure}
	\centering
	\includegraphics[width=0.46\columnwidth]{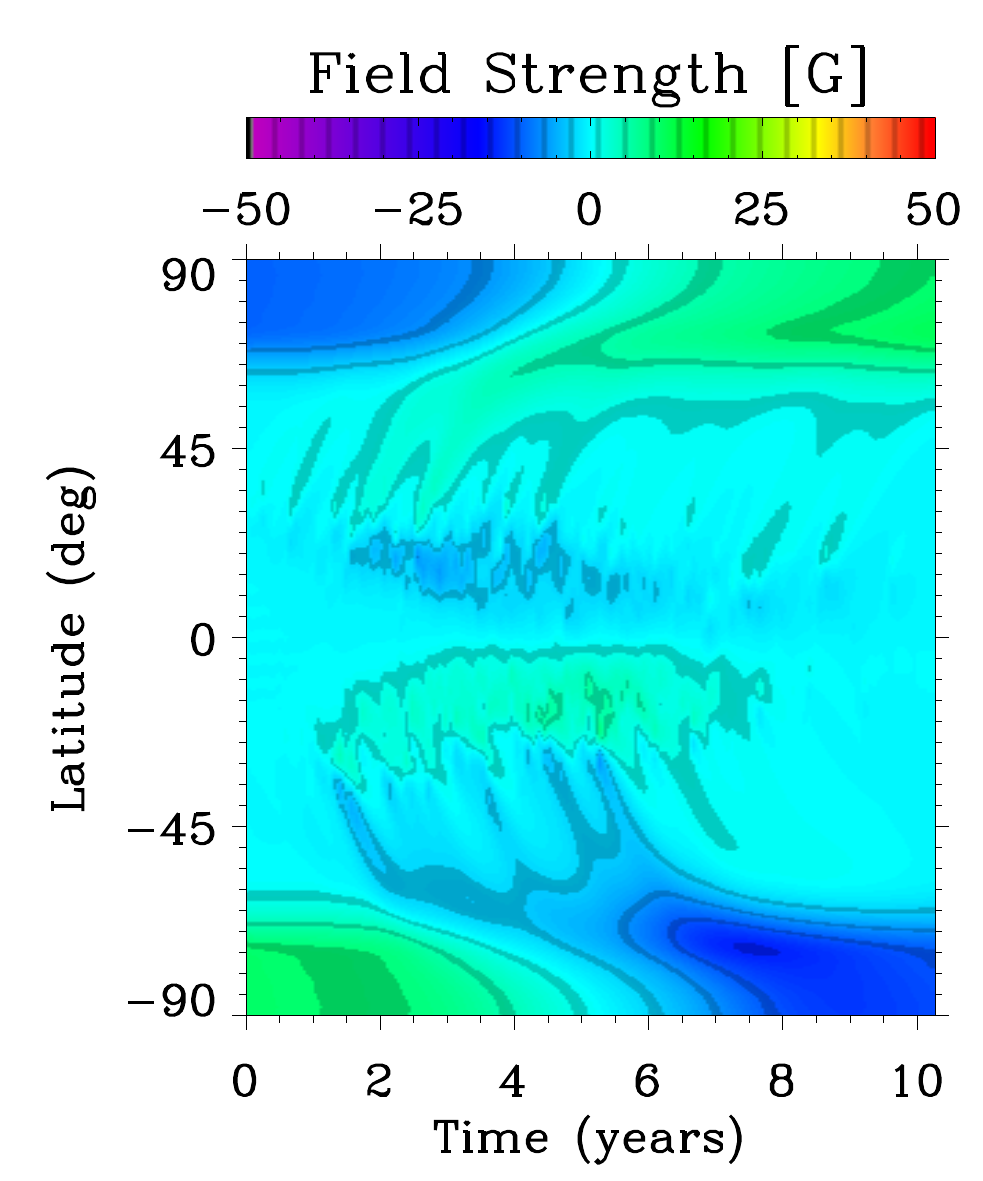}
	\includegraphics[width=0.46\columnwidth]{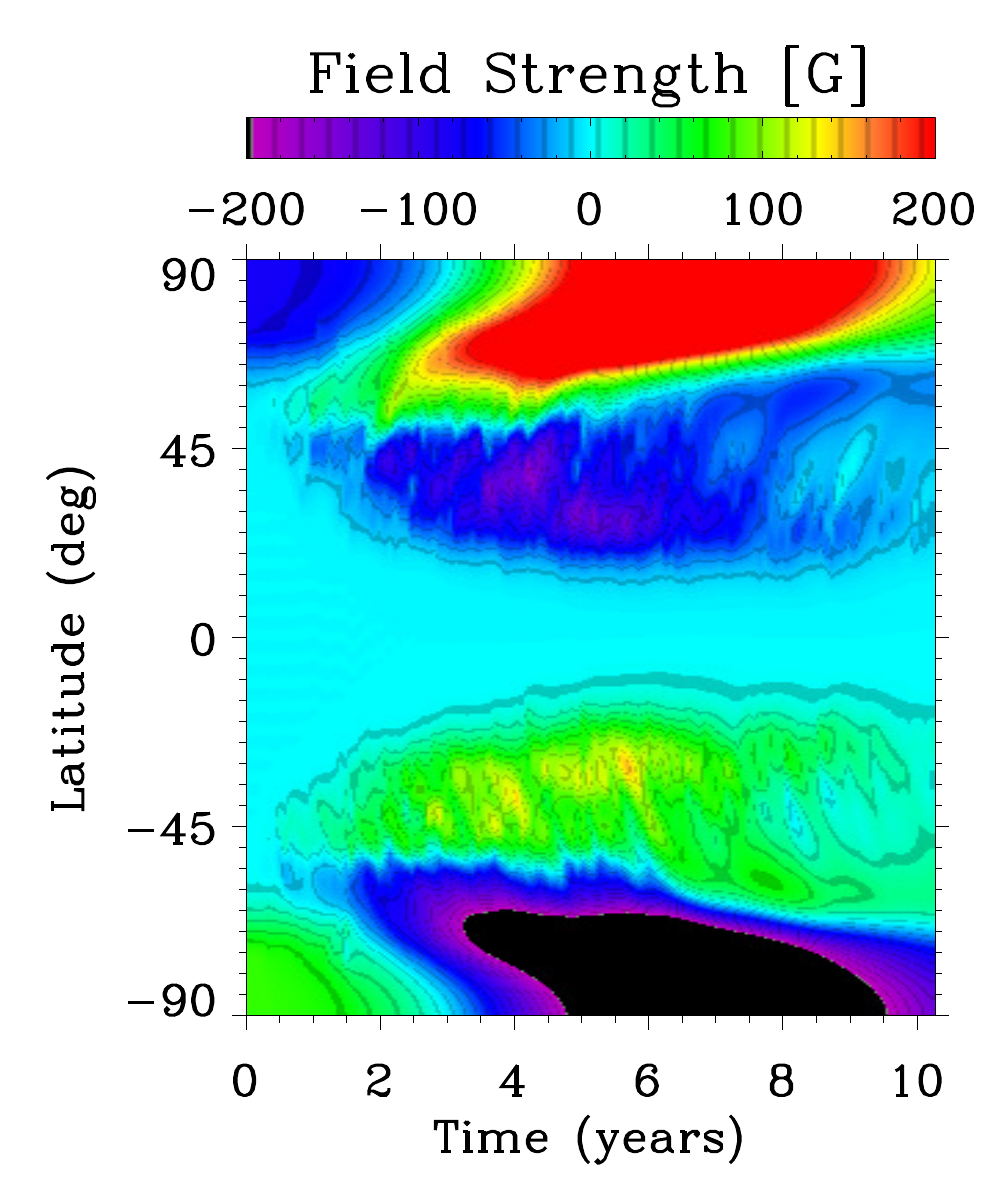}
	\includegraphics[width=0.46\columnwidth]{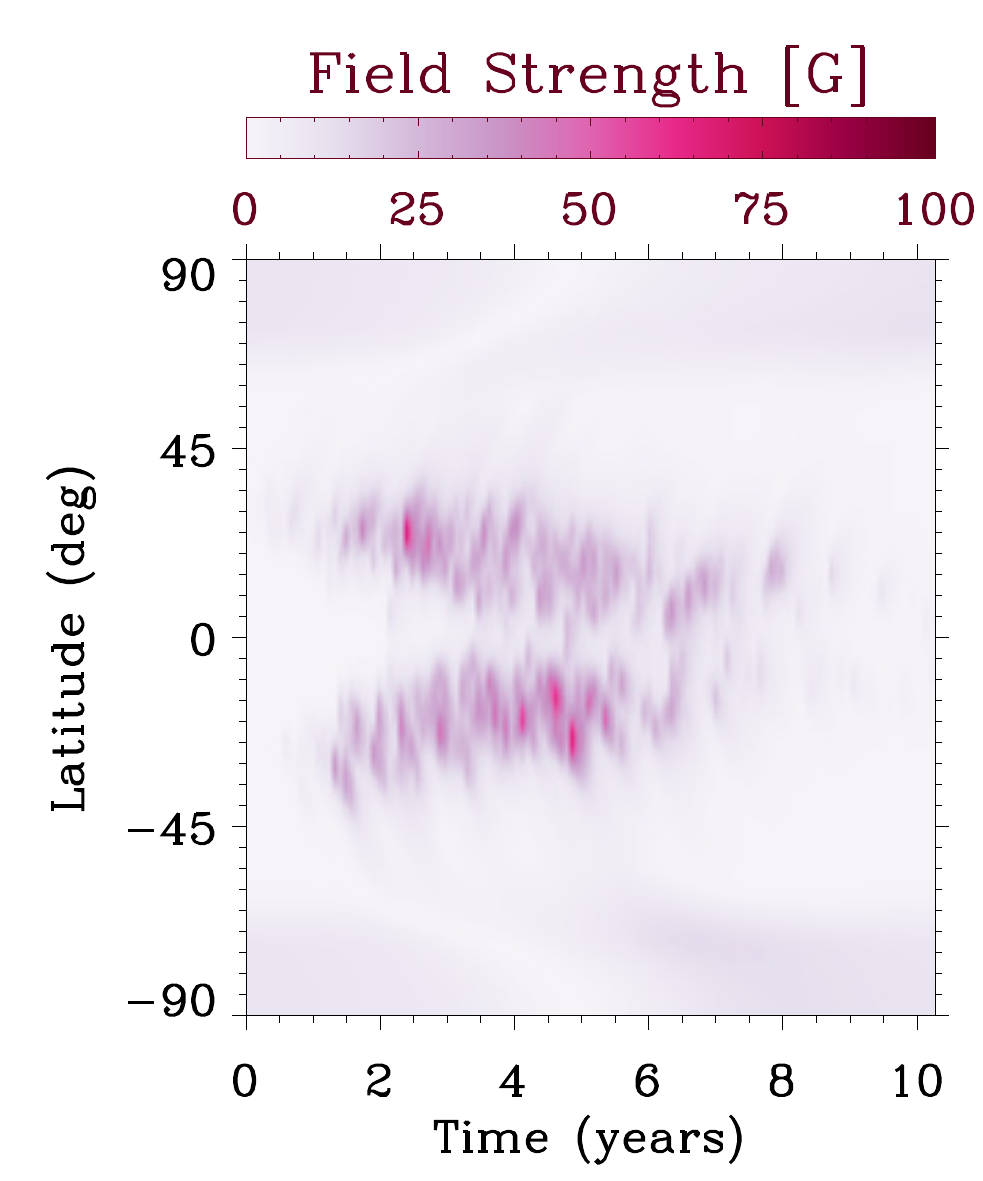}
	\includegraphics[width=0.46\columnwidth]{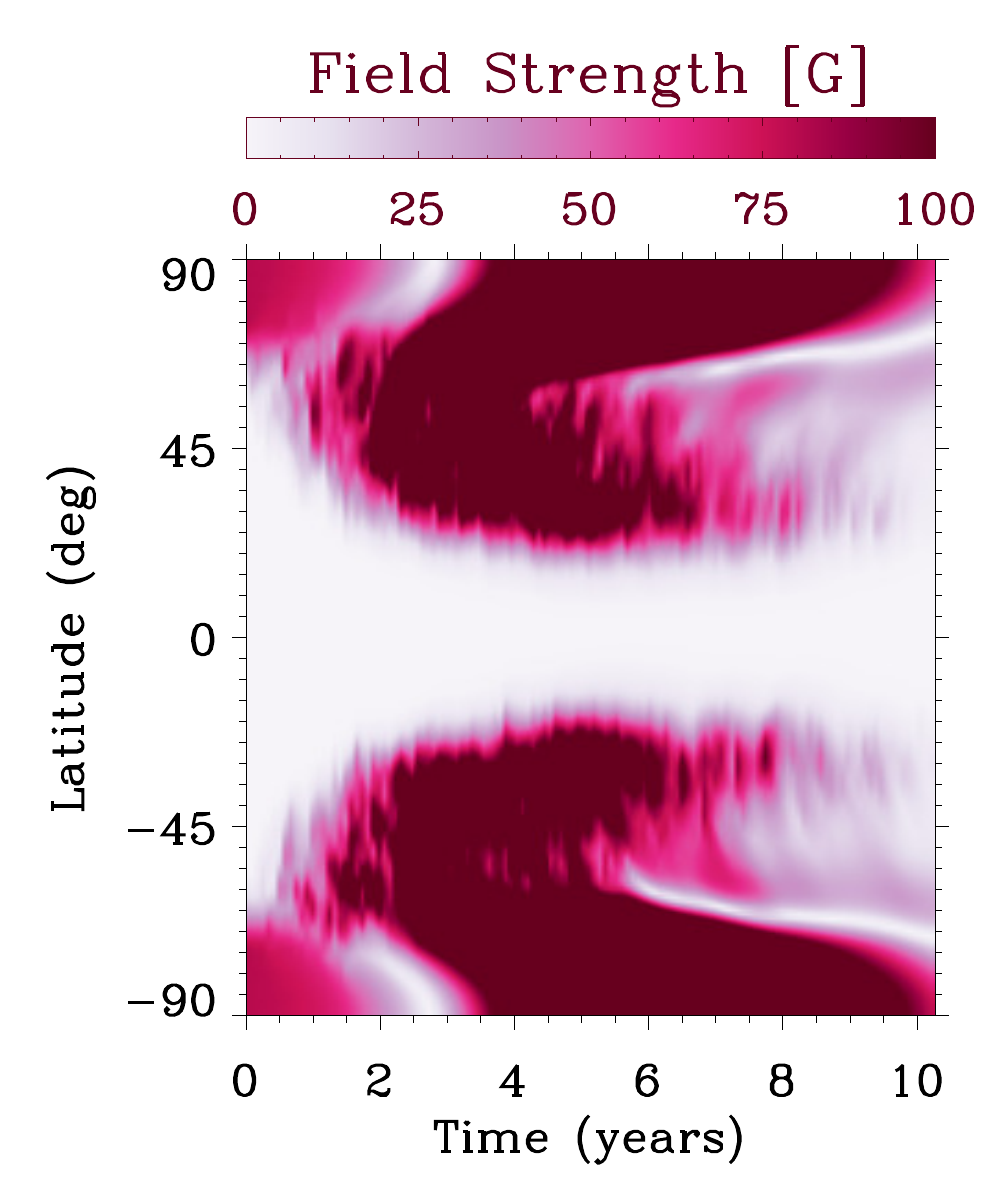}
	\caption{Azimuthal averages of signed (upper 
	panels) and unsigned (lower panels) field strength, as a function 
	of time, for solar (left) and eight times solar (right) rotation 
	rate. 	}
	\label{fig:magbfly}
\end{figure}

For the SFT parameters, we adopt values which were calibrated for simulations 
of solar activity for observed solar cycles \citep[][]{jcss11b}. The resulting 
longitudinally averaged signed and unsigned magnetic fields are shown in 
Fig.~\ref{fig:magbfly}, for the solar case and eight times more rapid rotation (also 
eight times more frequent emergence). For the latter case, 
the polar fields attain strengths comparable with the lower-latitude fields. 
The activity belts are also shifted towards higher latitudes, owing to increasing 
poleward deflection of rising flux tubes by the Coriolis effect. 
Formation of strong polar caps are also affected by increasing tilt angles of 
emerging bipolar regions, hence increasing dipole-moment contributions to the 
global (axial) dipole. 

\section{Modelling spot coverages}
As a first approximation to estimate starspot distributions, we set a threshold field 
strength, above which every pixel in the simulated magnetic map is treated as 
belonging to starspots. We 
determined the threshold value by requiring that the cycle-averaged global 
sunspot area coverage is about 0.2\%, representing a moderate to strong activity 
cycle. 
We then applied this criterion to our set of models with rotation rates 1 to 8 times 
that of the Sun. 

\begin{figure}
	\centering
	\includegraphics[width=0.8\columnwidth]{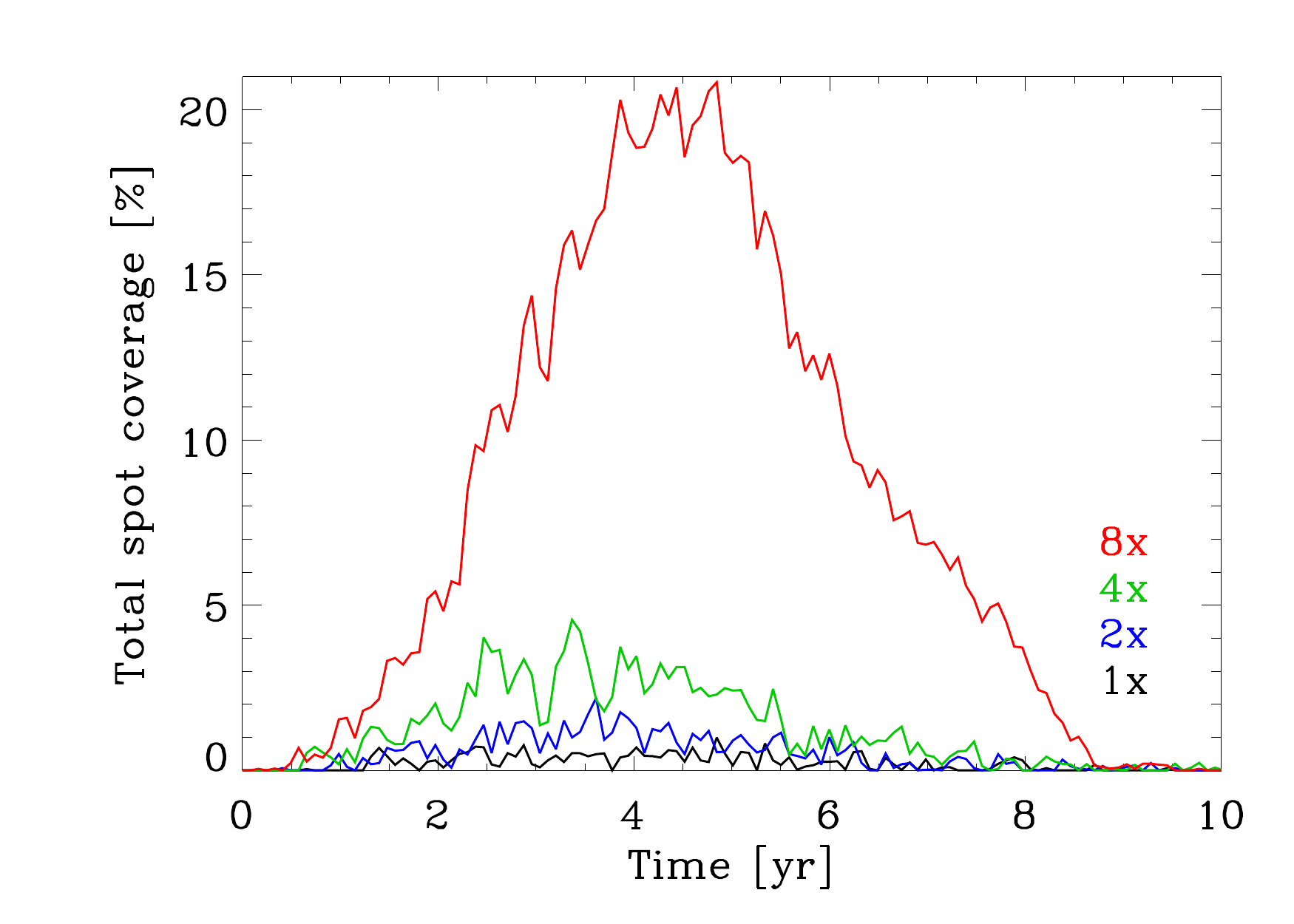}
	\caption{Cycle variation of the area covered by spots, for 1 to 8 
	times the solar rotation rate. 
	 	}
	\label{fig:area}
\end{figure}

Figure~\ref{fig:area} shows the evolution of the global spot coverage in the course of 
the activity cycle for given rotation rate (note that the activity level was assumed to 
scale linearly with the rotation rate). Our models correspond to sidereal rotation 
periods of about 
25, 12, 6, and 3 days, for which we assumed 1, 2, 4, and 8 times more frequent 
emergence than on the Sun. The 
area distribution of bipolar regions was kept unchanged. Among the models, 
the `8x' case deviates 
substantially from others, reaching spot coverages comparable to the lowest 
disc coverages observed for the G1.5V-type star EK Draconis, which rotates 
9 times faster than the Sun \citep{oneal04}. 

\section{Effect of nests of activity}
It is known that sunspot groups tend to emerge near sites of recent flux emergence. 
Statistics of such active nests were thoroughly investigated for the Sun 
\citep[e.g.,][]{castenmiller86,pelt10}. Our knowledge of the degree of nesting on other 
stars is currently limited by the low spatial resolution encountered in observational 
reconstructions of stellar surfaces. In the next stages of this project, 
we plan to reconstruct disc-integrated brightness variations out of our simulations. 
Such forward modelling can be useful in estimating the degree of starspot nesting 
from observations. 

\begin{figure}
	\centering
	\includegraphics[width=0.8\columnwidth]{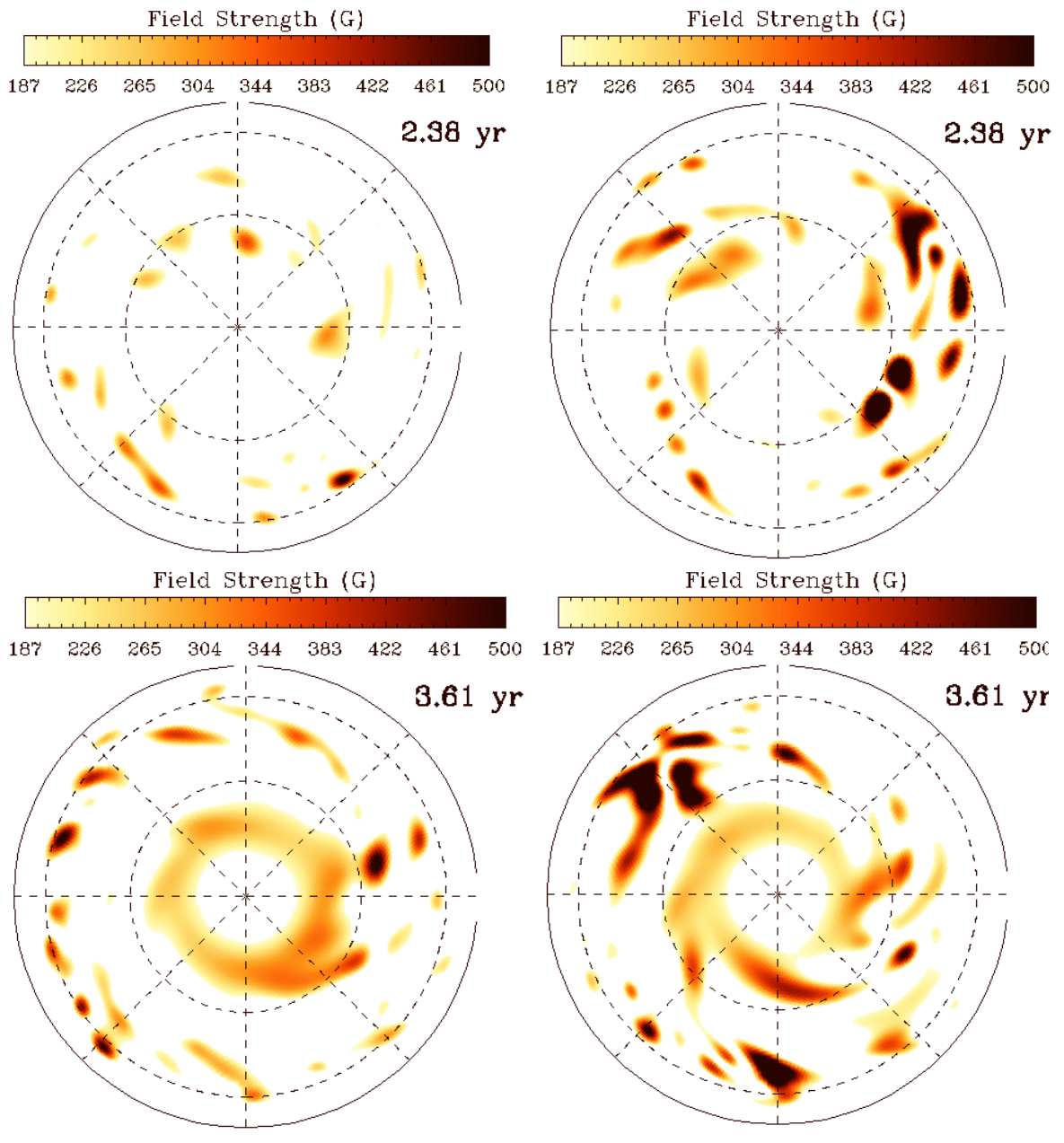}
	\caption{Pole-on snapshots of starspot distributions for a Sun-like star rotating 8 times faster than the Sun. Left: no nesting, Right: with a nesting probability of 70\%. 
	 	}
	\label{fig:nest}
\end{figure}

We included a nesting feature in the model by 
setting a probability for a given emergence to occur nearby recent emergence 
events. We assumed the probability that a given 
emerging bipolar region becomes a nest centre or that it belongs to a pre-existing 
nest to be 70\%. Figure~\ref{fig:nest} shows pole-on views from the `8x' 
simulations, with and without nesting. 
By visual inspection we can already assess 
that nesting can have a large impact on the longitudinal distribution of spots on active 
stars. The amplitude of rotational brightness variations would be significantly 
larger in the nested case as compared to a spot distribution with random longitudes. 
In addition, 
photometric detection of starspots would also be easier when they are nested, 
because smaller spots that are individually below a given detection threshold 
can be more easily detected when they are clustered, i.e., when they are more 
or less in phase and form a stronger photometric signal. 

\section{Summary and outlook}
We developed a model that calculates the surface distribution of radial magnetic 
field and starspots on Sun-like stars for a range of rotation periods between that 
of the Sun down to 3 days. The model assumes a one-to-one relation between the 
rotation rate and the flux emergence frequency, which is roughly consistent with 
empirical relations. This is in fact a free parameter that can be adjusted according to 
a desired observational relation. 

Our results suggest that polar spots should start to form close to about 
$P_{\rm rot}$=3 days. In our models, this happens due to the enhanced flux emergence 
rate and increased tilt angles. Consequently, the `polar spot' is largely unipolar, 
but it is surrounded by the opposite-polarity flux at high latitudes, where starspots 
can form. 

We plan to extend our models to multiple cycles with varying degrees of overlap, 
tilt angle scatter, and various possible perturbations of large-scale flows. 
To better describe the various variability patterns observed by the \emph{Kepler} 
mission, we will also include $(i)$ a more physical algorithm to decompose the 
emerging magnetic field into starspots and faculae, as well as $(ii)$ activity nesting 
as a function of the stellar activity level. 

\section*{Acknowledgments}
{EI acknowledges support by the Young Scientist Award Programme BAGEP-2016 
of the Science Academy, Turkey. This work has been partially supported by the BK21 
plus program through the National Research Foundation (NRF) funded by the Ministry 
of Education of Korea. AS acknowledges funding from the European Research Council 
under the European Union Horizon 2020 research and innovation 
programme (grant agreement No. 715947). }

\bibliographystyle{cs20proc}
\bibliography{Isik-CS20.bib}

\begin{thebibliography}{7}
\providecommand{\natexlab}[1]{#1}

\bibitem[\protect\astroncite{{Castenmiller}
  \emph{et~al.}}{1986}]{castenmiller86}
{Castenmiller}, M.~J.~M., {Zwaan}, C., \& {van der Zalm}, E.~B.~J. 1986,
  \solphys, 105, 237.

\bibitem[\protect\astroncite{{I{\c s}{\i}k} \emph{et~al.}}{2011}]{isik11}
{I{\c s}{\i}k}, E., {Schmitt}, D., \& {Sch{\"u}ssler}, M. 2011, \aap, 528,
  A135.

\bibitem[\protect\astroncite{{I{\c s}{\i}k} \emph{et~al.}}{2018}]{isik18}
{I{\c s}{\i}k}, E., {Solanki}, S.~K., {Krivova}, N.~A., \& {Shapiro}, A.~I.
  2018, \aap, 620, A177.

\bibitem[\protect\astroncite{{Jiang}
  \emph{et~al.}}{2011{\natexlab{a}}}]{jcss11a}
{Jiang}, J., {Cameron}, R.~H., {Schmitt}, D., \& {Sch{\"u}ssler}, M.
  2011{\natexlab{a}}, \aap, 528, A82.

\bibitem[\protect\astroncite{{Jiang}
  \emph{et~al.}}{2011{\natexlab{b}}}]{jcss11b}
{Jiang}, J., {Cameron}, R.~H., {Schmitt}, D., \& {Sch{\"u}ssler}, M.
  2011{\natexlab{b}}, \aap, 528, A83.

\bibitem[\protect\astroncite{{O'Neal} \emph{et~al.}}{2004}]{oneal04}
{O'Neal}, D., {Neff}, J.~E., {Saar}, S.~H., \& {Cuntz}, M. 2004, \aj, 128,
  1802.

\bibitem[\protect\astroncite{{Pelt} \emph{et~al.}}{2010}]{pelt10}
{Pelt}, J., {Korpi}, M.~J., \& {Tuominen}, I. 2010, \aap, 513, A48.

\end{thebibliography}

\end{document}